# MODELLING OPEN-SOURCE SOFTWARE RELIABILITY INCORPORATING SWARM INTELLIGENCE-BASED TECHNIQUES

Omar Shatnawi

Department of Computer Science, Al al-Bayt University, Mafraq, Jordan

## ABSTRACT

*In the software industry, two software engineering development best practices coexist: open-source and closed-source software. The former has a shared code that anyone can contribute, whereas the latter has a proprietary code that only the owner can access. Software reliability is crucial in the industry when a new product or update is released. Applying meta-heuristic optimization algorithms for closed-source software reliability prediction has produced significant and accurate results. Now, open-source software dominates the landscape of cloud-based systems. Therefore, providing results on open- source software reliability - as a quality indicator - would greatly help solve the open-source software reliability growth- modelling problem. The reliability is predicted by estimating the parameters of the software reliability models. As software reliability models are inherently nonlinear, traditional approaches make estimating the appropriate parameters difficult and ineffective. Consequently, software reliability models necessitate a high- quality parameter estimation technique. These objectives dictate the exploration of potential applications of meta- heuristic swarm intelligence optimization algorithms for optimizing the parameter estimation of nonhomogeneous Poisson process-based open-source software reliability modelling. The optimization algorithms are firefly, social spider, artificial bee colony, grey wolf, particle swarm, moth flame, and whale. The applicability and performance evaluation of the optimization modelling approach is demonstrated through two real open-source software reliability datasets. The results are promising.*

## 1. INTRODUCTION

The open-source software (abbreviated as OSS) and closed-source software are two best practices in the software industry. OSS is publicly available and allows modification without cost, whereas proprietary software is privately owned and developed by a closed team. The usage of OSS is growing rapidly, owing to its economic feasibility and security. Eighty to ninety percent of commercial applications contain OSS components [1]. The industry is transitioning from a closed source to a concurrently distributed environment in the OSS era. OSS is available all around us, and we use it and work with it daily. Operating systems (FreeBSD, Linux, and Solaris), database middleware technologies such as MySQL, Apache Web server, and even web browsers such as Mozilla Firefox are among them [2]. Developing reliable OSS products of increasing size and complexity is challenging. Mathematical modelling based on stochastic/statistical theories helps to represent and explain the OSS debugging phenomenon and quantitatively evaluate its reliability. Software-reliability models based on the non-homogeneous Poisson process (abbreviated as NHPP) have proven to be effective tools in practical software reliability engineering. They aid in describing the debugging process by providing trends, such as reliability growth and defect content [3-7].

Nature has, in many ways, inspired many researchers and is an essential source of inspiration. Accordingly, the majority of new algorithms are nature-inspired. Bioinspired, chemistry-, and physics-based algorithms are examples of nature-inspired optimization algorithms. Bio-inspired computing optimization algorithms, which are based on ideas and inspiration from biological evolution, are driving the development of new and robust competing techniques. Swarm intelligence- and evolutionary-based algorithms are examples of bioinspired algorithms [8, 9].

In nearly every science, engineering, and industry field, swarm-intelligence-based algorithms have been widely employed in real-world applications to solve nonlinear design problems [10]. One such application is parameter optimization. Optimization involves determining the best values of variables by minimizing or maximizing an objective function under specific constraints[11]. This involves identifying the parameters, creating an objective function, setting constraints, and employing an appropriate optimizer. A good swarm intelligence optimization algorithm, should continuously push search agents toward the global optimum and explore and exploit the defined search space [12].

Numerous NHPP-based software reliability models have been developed to estimate the reliability of OSS [13-18]. These models view debugging as a counting process characterized by their mean value functions (abbreviated as MVFs). Once the MVF is determined, the software reliability can be estimated. The model parameters are commonly estimated using maximum-likelihood estimation (abbreviated as MLE) or least-squares estimation (abbreviated as LSE). The optimization of these parameters is crucial. Although these two techniques are suitable for linear problems, most software reliability models are nonlinear. Therefore, they are inappropriate for estimating the parameters of the software reliability model because they are inadequate for solving local optimization problems [19, 20].

The nature-inspired approach is one of the approaches to overcome these limitations for parameter estimation of NHPP-based software reliability models [21-24]. However, most efforts have been made to model the debugging phenomena of close-source software. Therefore, this study explores the application of swarm intelligence-based algorithms, that is, the optimization modelling approach, to optimize the parameter estimation of an eminent NHPP-based software reliability model, namely, Goel-Okumoto model [3], which has been extended to develop software reliability models as a basic framework.

To the best of our knowledge, this study is the first to investigate the application of swarm intelligence-based algorithms to address the OSS reliability growth modelling optimization problem. The remainder of this paper is organized as follows. First, we briefly introduce swarm intelligence-based algorithms and an eminent NHPP-based software reliability model, namely, the Goel-Okumoto model. We then provide a criterion for validating and evaluating the optimization modelling approach. Subsequently, we validate and compare the optimization modelling approach with traditional approaches based on their descriptive performance. We then employed a k-fold cross-validation procedure to validate the predictive capability of the optimization modelling approach. Finally, the concluding remarks are presented.

## 2. SWARM INTELLIGENCE COMPUTING OPTIMIZATION ALGORITHMS

A swarm is a group of decentralized agents working together to achieve specific objectives. Swarm intelligence is a vital concept in artificial intelligence, which utilizes simple agents with minimal rules for emergent global behavior. Inspired by natural social swarms, swarm intelligence-based algorithms can model their behavior and provide robust, efficient, and economical solutions to optimization problems. These nature-inspired algorithms enable agents to find food sources and explore optimal routes despite their limited intelligence and skills [25].

Wang and Beni (1989) introduced swarm intelligence for cellular robotic systems [26]. The notion was first used in artificial intelligence research before it was applied to biological systems. In recent years, significant efforts have been made to adapt swarm intelligence-based techniques to various engineering problems. They typically share information among numerous agents, and multiple agents can be easily parallelized, making large-scale optimization more feasible from an implementation perspective [8]. Most swarm intelligence-based algorithms have a similar structure although they are defined in various forms. Overall, swarm intelligence leverages the principles of decentralized, self-organized, and emergent behavior to solve complex problems and efficiently perform tasks in a distributed manner.

Milon as (1993) defines the swarm intelligence paradigm as comprising five important concepts [27]. The first concept is *proximity*, which implies that the swarm should be capable of performing the necessary space/time computations. The second concept is *quality*, which implies that a swarm should be capable of adapting to environmental quality parameters. The third concept is the *diverse response*, which states that a swarm should not operate in overly restricted channels. The fourth concept is *stability*, which says that a swarm's behavior should not change as the environment changes. The fifth concept is *adaptability*, which means that a swarm should be able to alter the behaviour of the mote when the computational cost is justified. A typical flow chart of the swarm intelligence optimization techniques is depicted in Figure 1 to better understand how an optimization algorithm works. Initially, the population and related parameters should be initialized. Subsequently, the population's fitness values are computed in each iteration, and if the global best solution meets the termination criteria, swarm intelligence outputs the results [25, 28].

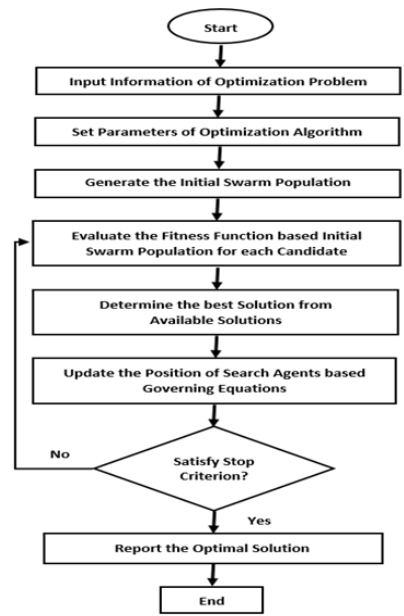

Figure 1. A typical flowchart of swarm intelligence computing optimization algorithms.

Numerous swarm intelligence-based algorithms have appeared over the last 30 years, inspired by animals and mammals, insects, birds, and fish and sea creatures that serve as inspiration [24, 29].

International Journal of Computer Science & Information Technology (IJCSIT) Vol 15, No 6, December 2023In this study, we employ the following algorithms for optimizing the parameter estimation of NHPP-based open-source software reliability modelling:

- Kennedy and Eberhart (1995) proposed *Particle Swarm Optimization* (abbreviated as PSO), which included a fundamental mechanism to replicate bird hunting behaviors [30]. It uses only the objective function and has few hyperparameters, focusing on finding optimal solutions in the solution spaces.

- Karaboga (2005) proposed the *Artificial Bee Colony* (abbreviated as ABC) by mimicking bee colonies' honey-gathering behaviors [31]. This involves three types of bees: employee bees collecting food from a specific source, onlooker bees patrolling the area, and scout bees searching for new locations. Food sources are defined as positions in the search space, with the quality of a food source determined by the fitness value of the objective function.

- Yang (2008) proposed the *Firefly Algorithm* (abbreviated as FA) by manipulating the behavior of fireflies [32]. It is unisex, with attractiveness being proportional to brightness. As the distance increases, less bright fireflies move towards brighter ones, whereas brighter ones move randomly. The landscape of the objective function determines firefly brightness.

- Mirjalili et al. (2014) introduced *Grey Wolf Optimization* (abbreviated as GWO), which stems from the modelling of grey wolf group predation behaviors and achieves optimization via the process of wolf pack tracking, surrounding, following, and attacking prey [33]. It seeks the optimal solution only when the best feasible solution falls within the optimal solution territory.

- Cuevas and Cienfuegos (2014) created *Social Spider Optimization* (abbreviated as SSO) by modelling social spider activities such as reciprocal predation, information exchange, reproduction, and progeny generations [34]. It modes a communal spider web, with every solution defining a spider and a weight determined by its fitness value. It mimics cooperative behavior in the colonies.

- Mirjalili (2015) proposed *Moth-Flame Optimization* (abbreviated as MFO), which was inspired by the transverse direction of moths toward the light source and is a realistic solution to tackling global optimization issues [35]. It uses a transverse navigation mechanism that relies on moonlight for night flights.

- Mirjalili and Lewis (2016) proposed the *Whale Optimization Algorithm* (abbreviated as WOA) for handling complicated optimization problems based on humpback whales' bubble-net hunting maneuver strategy [36]. It mimics humpback whale hunting behavior.

In the next section, we employ the aforementioned swarm intelligence-based algorithms for optimizing the parameter estimation of the Goel-Okumoto model. Owing to its simplicity, the model is still in use.

The NHPP treats the defect debugging process, that is, models the debugged defects time $t$ as a pure-birth counting process $(N_t, t \geq 0)$ with intensity function $\lambda_t$, for all $t \geq 0$, subject to [4, 7, 37]:

- $N_{t=0} = 0$ with a probability of 1.
- $(N_t, t \geq 0)$ exhibited independent increments.
- The probability that a defect debugs during $(t, t + \Delta t)$ is $\lambda_t \Delta t + 0(\Delta t)$.
- The probability that more than one defect will debug during $(t, t + \Delta t)$ is 0.

16

Accordingly, the NHPP-based software-reliability model, that is, the probability $N_t$ is a given integer $K(\geq 0)$ can $N_t$ be represented as follows:

$$Pr[\![N_t = k]\!] = \frac{(m_t)^k}{k!} \cdot \exp(-m_t) \qquad (1)$$

here $N_t$ be the expected value number of defects whose mean value function (abbreviated as MVF) is known as $m_t$.

Goel and Okumoto [3] developed the first NHPP model to represent the defect debugging phenomenon, which is a very simple form of $m_t$, based on the assumptions listed below:
- The software defect debugging phenomenon follows anNHPP with $m_t$.
- A software system is subject to failure owing to defects present in the system.
- Upon failure, the defect causing that failure is instantaneously debugged, and no other defects are introduced during the process.
- The expected number of defects debugged in $(t, t + \Delta t)$ is proportional to the number of defects that remain to be debugged.

Accordingly, the defect intensity, $\lambda_t$, at time $t$, that is, the Goel-Okumoto model, can be summarized in the following differential equation:

$$\lambda_t = \frac{\partial m_t}{\partial t} = b \cdot (a - m_t) \qquad (2)$$

here, $m_t$ is the cumulative number of defects debugged at a certain time $t$, $a$ is the defect-content, and $b$ is the proportionality constant that represents the defect debugged rate per defect. Solving Eq. (2) with the initial-condition $m_{t=0} = 0$, yields its MVF which shows an exponential-growth curve given by

$$m_t = a \cdot (1 - \exp(-b \cdot t)) \qquad (3)$$

## 3. EXPERIMENTAL EVALUATION

### a. Experimental Setup and Evaluation Metrics

Software-reliability trend analysis is used to assess the progress of the OSS debugging process. As a result, before employing the modelling approach, it is plausible to decide whether the OSS reliability dataset exhibits software reliability growth. For this reason, two trend test techniques of usual practice, the Laplace factor and arithmetic average [5, 6, 37, 38], are utilized and given

$$Laplace\ Factor = \frac{\sum_{i=1}^{k}(i-1)t_i - \frac{k-1}{2}\sum_{i=1}^{k}t_i}{\sqrt{\frac{k^2-1}{12}\sum_{i=1}^{k}t_i}} \qquad (4)$$

$$Arithmetic\ Average = \frac{1}{k}\sum_{i=1}^{k}t_i \qquad (5)$$

here, $k$ is the number of defects and $t_i$ is the time of occurrence of failure $i(i = 1, ..., k)$.
For the Laplace factor (arithmetic average), negative (decreasing) values promoted reliability growth.

We chose Apache project-based open-source experimental software reliability datasets for our analysis because they are used mainly in defect prediction studies [39-41]. The first dataset (Apache 2.0.36) and the second dataset (Apache 2.0.39) were tested and debugged over 103 (164) days, and in the end, 50 (58) defects were detected by the debugging team. Table 1 tabulated the open-source datasets in terms of data points $(t_i, y_i)$, $(i = 0, 1, 2, ..., 103\ (164))$ where $y_i$ where is the cumulative number of defects debugged by $t_i(0 < t_1 < t_2 < \cdots <$ $t_{103}(t_{164})$). To avoid repetition, certain defect data are not shown. For example, because no defects are detected on the sixth day, no defect data is shown on this day for the first dataset.

To measure the performance of the optimization modelling approach, its ability to fit past data (descriptive-performance) and predict future behaveior adequately from present/past data (predictive-capability) are examined [4, 37, 42-47]. There are several vital goodness-of-fit statistics for regression analysis. Our analysis examines the same goodness-of-fit test measure, namely, the sum of squared error (abbreviated as SSE) across traditional regression procedures (e.g., MLE) and swarms' intelligence-based algorithms inspired by animals (e.g., GWO), insects (e.g., ABC, SSO, FA, and MOF), birds (e.g., PSO and CS), and sea creatures (e.g., WOA) to optimize parameter estimation for the Goel-Okumoto model, thus allowing a fair comparison. Assessing numeric measures of goodness-of-fit, SSE, using

$$SSE = \sum_{i=1}^{f}(\widehat{m_{t_i}} - y_i) \qquad (6)$$

SSE assesses the dispersion between estimated values $\widehat{m_{t_i}}$ and the actual data $y_i$ of the dependent variable. The lower the value of SSE, the better the approach fit.

Table 1. Datasets used for experimentation.

| Apache 2.0.3h6 | | | | Apache 2.0.39 | | | |
|---|---|---|---|---|---|---|---|
| $t_i$ | $y_i$ | $t_i$ | $y_i$ | $t_i$ | $y_i$ | $t_i$ | $y_i$ |
| 1 | 2 | 31 | 35 | 1 | 1 | 26 | 35 |
| 2 | 7 | 32 | 36 | 2 | 3 | 28 | 36 |
| 3 | 8 | 33 | 39 | 3 | 5 | 29 | 37 |
| 4 | 9 | 34 | 40 | 4 | 8 | 30 | 39 |
| 5 | 10 | 35 | 43 | 5 | 11 | 31 | 40 |
| 7 | 12 | 38 | 46 | 7 | 13 | 32 | 41 |
| 8 | 13 | 40 | 47 | 8 | 14 | 35 | 44 |
| 9 | 14 | 43 | 48 | 9 | 15 | 38 | 45 |
| 10 | 17 | 44 | 49 | 10 | 16 | 39 | 46 |
| 12 | 19 | 103 | 50 | 11 | 17 | 42 | 47 |
| 13 | 20 | | | 15 | 20 | 43 | 48 |
| 15 | 22 | | | 16 | 22 | 49 | 51 |
| 17 | 23 | | | 17 | 25 | 50 | 52 |
| 18 | 25 | | | 18 | 26 | 51 | 53 |
| 21 | 26 | | | 19 | 27 | 57 | 54 |
| 25 | 27 | | | 22 | 30 | 66 | 55 |
| 27 | 28 | | | 23 | 31 | 70 | 56 |
| 29 | 30 | | | 24 | 32 | 81 | 57 |
| 30 | 32 | | | 25 | 34 | 164 | 58 |

For the predictive performance comparison, we study the $k$-fold cross-validation technique, which splits the dataset into $k$ groups to validate the optimization modelling approach on one group while training it on the remaining $k - 1$ groups to obtain a satisfactory generalization ability, all of which $k$ times. As a result, we explore two widely used k values when $k = 2$ and $k = 10$ [48].

All the experiments are conducted using a computer with a 12th Gen Intel(R) Core(TM) i7 − 1255U CPU running at 1.70 GHz with 16.0 GB of RAM and a 64−bit Windows 11Pro architectural, x64−based microprocessor. Their source code is implemented using MATLAB for the seven swarm intelligence-based optimization techniques (R2021a).

For a fair comparison, the parameter settings of all algorithms in terms of the population size, number of iterations, and limits of the input variables within which the optimizer is allowed to search are the same. However, the statistical package for social sciences (abbreviated SPSS) based on the linear regression method, has been used in traditional techniques.

### 3.2. EXPERIMENTAL RESULTS AND COMPARISONS

Both trend tests are used to determine whether the given OSS reliability dataset promotes reliability-growth. If it does, we estimate the parameters of the Goel-Okumoto model using both traditional and swarm intelligence approaches. The goodness-of-fit test is then applied. Based on the findings, we conduct a comparative evaluation of the two modelling approaches under consideration. Finally, we perform a test to determine the superior approach's predictive validity.

#### 3.2.1. First Dataset / Apache 2.0.36

Figures 2 and 3 show the first dataset/Apache 2.0.36 relevant results for the Laplace factor and arithmetic average trend tests, respectively. Except for the first two days, the values are negative (decreasing), indicating increased reliability. Initially, their values indicate decreased reliability, which is generally expected. Then, after the second day of commencement, their values increased indicating reliability growth, which is usually welcomed after reliability decreases. As a result, the first dataset/Apache 2.0.36 is appropriate for software-reliability modelling.

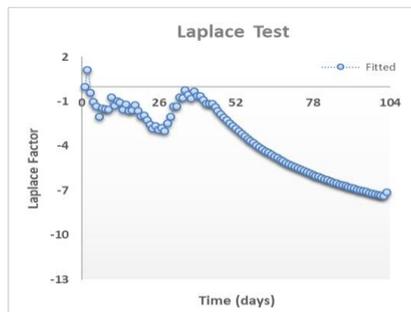
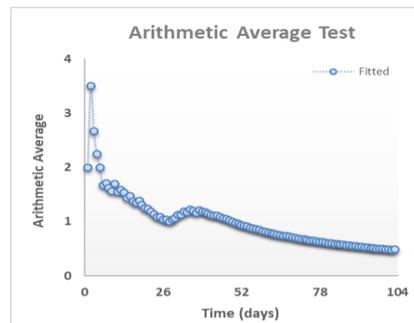

Figure 2. Results of the dataset's Laplace factor test.

Figure 3. Results of the dataset's arithmetic average test.

Table 2 summarizes the estimated values of the undetermined parameters and the goodness-of-fit test results for the two approaches based on the Goel-Okumoto model. It is clear that all other algorithms have reasonably comparable performance comparison results. The comparison results show that all seven swarm intelligence computing optimization algorithms (FA, SSO, ABC, GWO, PSO, MFO, and WOA) outperformed the other traditional estimation techniques (MLE).

Table 2. Parameter-estimation and comparison criterion.

| Techniques | Parameter Estimation | | Comparison Criterion |
|---|---|---|---|
| | *a* | *b* | *SSE* |
| MLE | 52.3160 | 0.0393371 | 917.8284 |
| FA | 52.2411 | 0.0394581 | 915.4488 |
| SSO | 52.2434 | 0.0394348 | 915.4543 |
| ABC | 52.2432 | 0.0394575 | 915.4484 |
| GWO | 52.2674 | 0.0394414 | 915.4726 |
| PSO | 52.2437 | 0.0394572 | 915.4484 |
| MFO | 52.2437 | 0.0394570 | 915.4484 |
| WOA | 52.2444 | 0.0394549 | 915.4484 |

Figure 4 shows the convergence curves of the seven swarm intelligence computing optimization algorithms based on the Goel-Okumoto model for the first dataset/Apache 2.0.36. The convergence curve for an algorithm shows the best obtained SSE values for 100 iterations. However, the total number of iterations was chosen to be 100, and most of the time, the best solution remained the same after 50 iterations. The fitting results of the noncumulative and cumulative first dataset/Apache 2.0.36 for the seven swarm intelligence computing optimization algorithms based on the Goel-Okumoto model are shown in Figures 5 and 6. It is clear that they fit the first dataset/Apache 2.0.36 reasonably well.

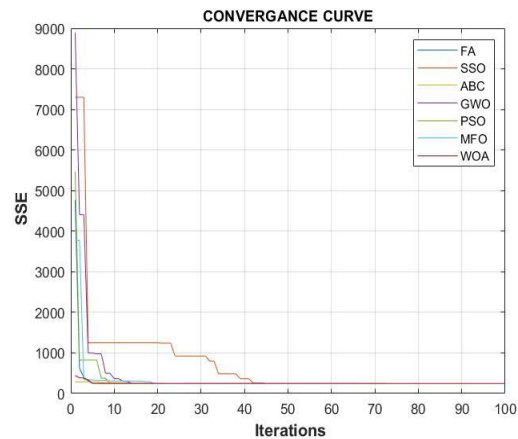

Figure 4. Convergence curve of all algorithms.

The swarm intelligence algorithms were evaluated using k−fold cross-validation, with k = 10 and k = 2. For each algorithm, the Goel-Okumotomodel parameters were optimized to maximize the overall performance in each case. The metric chosen for the analysis is SSE over the training/testing first dataset/Apache 2.0.36. Tables 3 and 4 give both 10− and 2−fold cross-validation results, respectively. It is clear that all algorithms have reasonably comparable performance comparison results for the first dataset/Apache 2.0.36. Based on the first dataset/Apache 2.0.36 analyses and approach comparisons, the optimization modelling.

Approach demonstrates strong descriptive and predictive power. From these results, we may conclude that the optimization modelling approach is a more useful model for software reliability measurement.

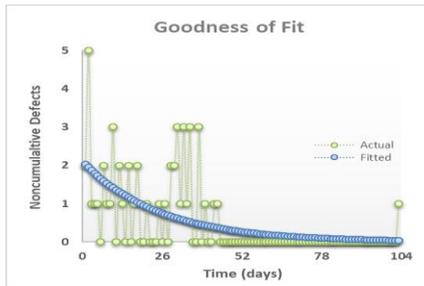 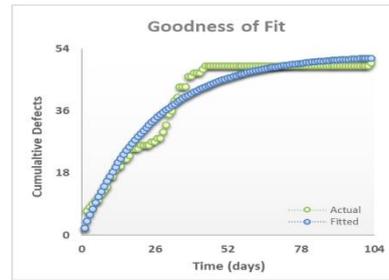

Figure 5. Results of the modelling approach's fitting of the noncumulative dataset.

Figure 6. Results of the modelling approach's fitting of the cumulative dataset

Table 3. Results of 10-fold cross-validation experiment.

| Techniques | Parameter Estimation | | 10-fold cross-validation | |
|---|---|---|---|---|
| | $a$ | $b$ | $SSE_{Training}$ | $SSE_{Testing}$ |
| FA | 52.1900 | 0.039659 | 911.0025 | 915.6235 |
| SSO | 51.7752 | 0.040744 | 915.7660 | 922.1435 |
| ABC | 52.2047 | 0.039641 | 911.0060 | 915.6160 |
| GWO | 52.1967 | 0.039602 | 911.0246 | 915.5330 |
| PSO | 52.1936 | 0.039653 | 911.0021 | 915.6181 |
| MFO | 52.1936 | 0.039653 | 911.0021 | 915.6183 |
| WOA | 52.1948 | 0.039650 | 911.0022 | 915.6132 |

Table 4. Results of 2-fold cross-validation experiment.

| Techniques | Parameter Estimation | | 2-fold cross-validation | |
|---|---|---|---|---|
| | $a$ | $b$ | $SSE_{Training}$ | $SSE_{Testing}$ |
| FA | 52.9780 | 0.038265 | 869.9547 | 928.2589 |
| SSO | 54.0406 | 0.036387 | 888.3470 | 985.7705 |
| ABC | 52.9929 | 0.038240 | 869.9620 | 928.7651 |
| GWO | 52.9460 | 0.038273 | 869.9798 | 926.9860 |
| PSO | 52.9720 | 0.038261 | 869.9517 | 927.9870 |
| MFO | 52.9724 | 0.038261 | 869.9517 | 927.9997 |
| WOA | 52.9769 | 0.038250 | 869.9521 | 928.1372 |

### 3.2.2. Second Dataset / Apache 2.0.39

Figures 7 and 8 show the second dataset/Apache 2.0.39 relevant results for the Laplace factor and arithmetic average trend tests, respectively. Except for the first five days, the values are negative (decreasing), indicating increased reliability. Initially, their values indicate decreased reliability, which is considered normal. Then, after the fifth day of commencement, their values increased indicating reliability growth, which is usually welcomed after reliability decreases. As a result, the second dataset/Apache 2.0.39 is appropriate for software-reliability modelling.

International Journal of Computer Science & Information Technology (IJCSIT) Vol 15, No 6, December 2023

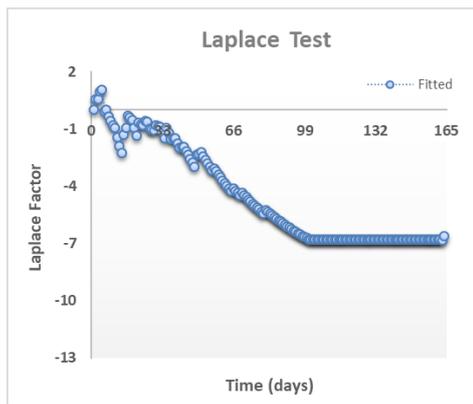
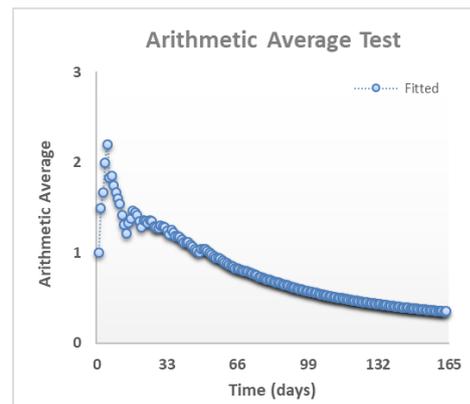

Figure 7. Results of the dataset's Laplace factor test.

Figure 8. Results of the dataset's arithmetic average test.

Table 5 summarizes the estimated values of the undetermined parameters and the goodness-of-fit test results for the two approaches based on the Goel-Okumoto model. It is clear that, except for SSO, all other algorithms have reasonably comparable performance comparison results. The comparison results show that all seven swarm intelligence computing optimization algorithms (FA, SSO, ABC, GWO, PSO, MFO, and WOA) outperformed the other traditional estimation techniques (MLE).

Table 5. Parameter-estimation and comparison criterion.

| Techniques | Parameter Estimation | | Comparison Criterion |
|---|---|---|---|
| | *a* | *b* | *SSE* |
| MLE | 58.3830 | 0.0367070 | 429.4830 |
| FA | 58.3715 | 0.0366210 | 389.8408 |
| SSO | 58.0075 | 0.0374998 | 400.0120 |
| ABC | 58.3701 | 0.0366121 | 389.8390 |
| GWO | 58.3988 | 0.0366241 | 389.9644 |
| PSO | 58.3703 | 0.0366116 | 389.8390 |
| MFO | 58.3704 | 0.0366117 | 389.8390 |
| WOA | 58.3702 | 0.0366122 | 389.8390 |

Figure 9 shows the convergence curves of the seven swarm intelligence computing optimization algorithms based on the Goel-Okumoto model for the second dataset/Apache 2.0.39. The convergence curve for an algorithm shows the best obtained SSE values for 100 iterations. However, the total number of iterations was chosen to be 100, and most of the time, the best solution remained the same after 50 iterations. The fitting results of the noncumulative and cumulative second dataset/Apache 2.0.39 for the seven swarm intelligence computing optimization algorithms based on the Goel-Okumoto model are shown in Figures 10 and 11. It is clear that they fit the second dataset/Apache 2.0.39 reasonably well.

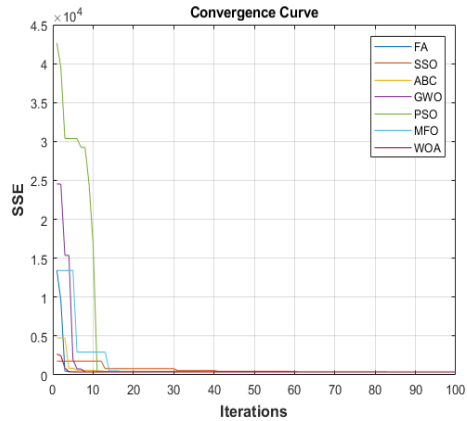

Figure 9. Convergence curve of all algorithms.

The swarm intelligence algorithms were evaluated using k−fold cross-validation, with k = 10 and k = 2. For each algorithm, the Goel-Okumoto model parameters were optimized to maximize the overall performance in each case. The metric chosen for the analysis is SSE over the training/testing second dataset/Apache 2.0.39. Tables 6 and 7 give both 10− and 2−fold cross-validation results, respectively. It is clear that all algorithms have reasonably comparable performance comparison results for the second dataset/Apache 2.0.39. Based on the second dataset/Apache 2.0.39 analyses and approach comparisons, the optimization modelling approach demonstrates strong descriptive and predictive power. From these results, we may conclude that the optimization modelling approach is a more useful model for software reliability measurement.

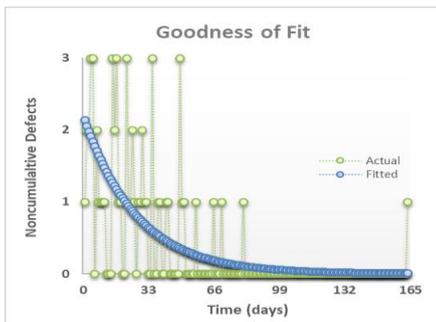

Figure 10. Results of the modelling approach's fitting of the noncumulative dataset.

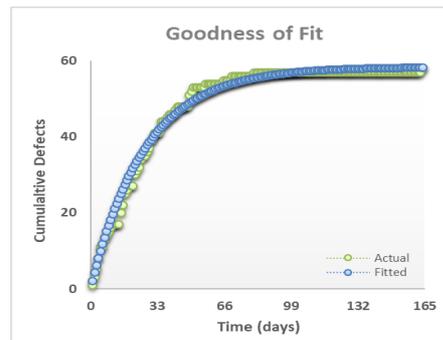

Figure 11. Results of the modelling approach's fitting of the cumulative dataset.

Table 6. Results of 10-fold cross-validation experiment.

| Techniques | Parameter Estimation | | 10-fold cross-validation | |
| --- | --- | --- | --- | --- |
| | $a$ | $b$ | $SSE_{Training}$ | $SSE_{Testing}$ |
| FA | 58.4189 | 0.036522 | 385.1514 | 389.9998 |
| SSO | 58.2708 | 0.036584 | 387.5057 | 391.2536 |
| ABC | 58.4278 | 0.036517 | 385.1443 | 390.0639 |
| GWO | 58.4558 | 0.036487 | 385.2003 | 390.3446 |
| PSO | 58.4278 | 0.036518 | 385.1443 | 390.0635 |
| MFO | 58.4278 | 0.036517 | 385.1443 | 390.0638 |
| WOA | 58.4281 | 0.036516 | 385.1443 | 390.0662 |

International Journal of Computer Science & Information Technology (IJCSIT) Vol 15, No 6, December 2023

Table 7. Results of 2-fold cross-validation experiment.

| Techniques | Parameter Estimation | | 2-fold cross-validation | |
|---|---|---|---|---|
| | $a$ | $b$ | $SSE_{Training}$ | $SSE_{Testing}$ |
| FA | 58.2748 | 0.036492 | 362.6554 | 391.9648 |
| SSO | 58.3618 | 0.036386 | 363.2166 | 390.9157 |
| ABC | 58.2703 | 0.036510 | 362.6519 | 391.9148 |
| GWO | 58.2616 | 0.036472 | 362.7060 | 392.6340 |
| PSO | 58.2705 | 0.036510 | 362.6520 | 391.9079 |
| MFO | 58.2703 | 0.036510 | 362.6519 | 391.9144 |
| WOA | 58.2704 | 0.036510 | 362.6520 | 391.9148 |

## 4. CONCLUSIONS

Software-reliability predictions are subject to much research, and various models and techniques exist. Scalable computing and artificial intelligence advance swarm intelligence, utilizing natural social organism behavior. This paper explores employing swarm intelligence-based algorithms to help solve the OSS reliability growth-modelling problem. The optimization modelling approach has been employed to optimize the parameter estimation of an eminent NHPP-based software reliability model, namely the Goel-Okumoto model.

Experimental results and comparisons reveal that employing swarm intelligence-based algorithms brings the results closer to the actual OSS debugging scenario. Therefore, these algorithms are robust and effective at resolving parameter optimization issues for software reliability modelling. The extension of the optimization modelling approach to address more realistic scenarios, such as the concept of change-point problem and imperfect-debugging phenomenon during the OSS debugging process, is an uphill task that inspires future studies.

**ACKNOWLEDGEMENTS**

The author acknowledges with gratitude the anonymous reviewers for their helpful and constructive comments**REFERENCES**

[1] Frank Nagle and Jenny Hoffman, (2020). *The Hidden Vulnerabilities of Open Source Software*, 2020. Retrieved January 23, 2023, from: http://hbswk.hbs.edu/item/the-hidden-vulnerabilities-of-open-source-software
[2] Shiva Tyagi, Devendra Kumar, and Sachin Kumar. (2019). Open source software: analysis of available reliability models keeping security in the forefront. *International Journal of Information Technology*, pages 1–10.
[3] Amrit L Goel and KazuOkumoto. (1979). Time-dependent error-detection rate model for software reliability and other performance measures. *IEEE transactions on Reliability*, 28(3):206–211.
[4] PK Kapur, Hoang Pham, Anshu Gupta, PC Jha, et al. (2011). *Software reliability assessment with OR applications*, volume 364. Springer.
[5] Omar Shatnawi. (2014). Measuring commercial software operational reliability: an interdisciplinary modelling approach. *EksploatacjaiNiezawodnos´c´*, 16(4).
[6] Omar Shatnawi. (2016). An integrated framework for developing discrete-time modelling in software reliability engineering. *Quality and Reliability Engineering International*, 32(8):2925–2943.
[7] Shigeru Yamada. (2014). *Software reliability modeling: fundamentals and applications*, volume 5. Springer.24